\documentclass[12pt]{iopart}
\usepackage{graphicx}

\begin{document}

\title
[Self-consistent Wigner distribution function study of gate-voltage
 controlled TBRTD]
{Self-consistent Wigner distribution function study of gate-voltage
 controlled triple-barrier resonant tunnelling diode}

\author
{P W\'ojcik$^1$, B J Spisak$^{1,2}$, M Wo{\l}oszyn$^1$ and J Adamowski$^1$}

\address
{$^1$ Faculty of Physics and Applied Computer Science,
AGH University of Science and Technology,
Al. Mickiewicza 30, 30-059 Krak\'ow, Poland}

\address
{$^2$ School of Physics and Astronomy, University of Leeds,
Leeds LS2~9JT, United Kingdom}

\ead{spisak@novell.ftj.agh.edu.pl}


\submitto{Semiconductor Science and Technology}

\maketitle

\begin {abstract}
The electron transport through the triple-barrier resonant
tunnelling diode (TBRTD) have been studied by the self-consistent
numerical method for the Wigner-Poisson problem.
The electron flow through the TBRTD can be
controlled by the gate voltage applied to one of the
potential well regions. For different gate voltage values we have
determined the current-voltage characteristics, potential energy
profiles, and electron density distribution. We have found the
enhancement of the peak-to-valley ratio (up to $\sim$10), the
appearance of the linear current versus bias voltage behaviour
within the negative-differential resistance region, and
the bistability of the current-voltage characteristics. The analysis
of the self-consistent potential energy profiles and electron
density distribution allowed us to provide a physical
interpretation of these properties.
\end{abstract}

\section{Introduction}\label{sec:intro}

The standard analysis of transport processes in solids
is based on the semiclassical Boltzmann
transport equation, e.g.~\cite{Baraff1997, Logan1991, Barnas1990, Sarker1987}.
However, in the solid-state nanostructures, the quantum effects become more
pronounced and the semiclassical description fails~\cite{{Wacker1999}}.
In particular, the theoretical description of electron transport in
semiconductor nanostructures and nanodevices requires the fully
quantum-mechanical treatment.
In the literature, one can find several quantum-mechanical approaches to the
electronic transport in nanodevices, such as those based on the envelope
function~\cite{Harrison_QWWD, Datta1999, Landauer1970}, density
matrix~\cite{Qi2007, Iotti2005, Bufler1994, Mizuta1991}, non-equilibrium Green
function~\cite{Knezevic2004, Cresti2003, Datta2000, Lake1997}, and non-classical
distribution function~\cite{Nedjalkov2004, Bertoni1999, Bordone1999}.
One of the most transparent and powerful methods, used to investigate the
quantum electron dynamics in the nanodevices, is based on the quantum kinetic
equation for the Wigner distribution function (WDF)~\cite{Wigner32, hosw84,
Lee95}.

The quantum kinetic equation for the WDF has been applied to a description of
the electron transport in the resonant tunnelling diode (RTD), i.e. the double
barrier semiconductor nanostructure~\cite{Nedjalkov2004, Kim2001, Mains1988,
Frensley1987, Kluksdahl1987}.
The triple-barrier resonant tunnelling diode (TBRTD)~\cite{Romo2002, Macks1996}
is a natural extension of double barrier RTD.
In the TBRTD, we deal with two quantum wells separated by the barrier regions.
The resonant electron tunnelling, which is responsible for the successful
operation of the resonant-tunnelling nanodevices, strongly depends on the
relative positions of energy levels of the electrons confined in the different
quantum-well regions.
Therefore, the TBRTD nanodevice provides a possibility of additional tuning of
the resonant condition by applying a gate voltage to the one of the quantum-well
regions.
The preliminary results for the TBRTD nanodevice are presented in our previous
paper~\cite{Wojcik2008}.
These results~\cite{Wojcik2008} have been obtained using the model that neglects
the electrostatic field produced by the electrons gathered in the nanodevice.
In the present paper, we study the electron transport through the gate-voltage
controlled TBRTD nanodevice taking into account this electrostatic field by a
self-consistent procedure.
For this purpose, we apply a numerical method of self-consistent solution of the
quantum kinetic equation for WDF and the Poisson equation.
As a main result, we obtain the current-voltage characteristics of the TBRTD as
functions of nanodevice parameters and gate potential.
This approach allows us to simulate the operation of the TBRTD nanodevice.

The present paper is organized as follows.
In Sec. \ref{sec:model} we introduce the quantum kinetic approach and describe
the model of the nanodevice.
In Sec. \ref{sec:results1} we present the results and their discussions, and in
Sec. \ref{sec:concl}, we provide the conclusions and summary.

\section{Theoretical Method and Model of Nanodevice}\label{sec:model}

The resonant-tunnelling devices consist of different semiconductor layers which
act as potential barriers and wells for the charge carriers.
The transport properties of RTD and TBRTD are determined by the chemical and
physical properties of the semiconductors forming the layers as well as by the
thicknesses of the layers.
At the semiconductor heterojunction, the difference in energy gaps between two
different materials leads to a conduction (valence) band discontinuity. The
position-dependent conduction band minimum forms
the potential profile for electrons with potential barrier and well regions.
In the potential well regions, the electrons can form the quasi-bound states.
These states strongly affect the electron transport via the nanostructure due to
the quantum interference effects which -- under certain conditions --
can lead to the resonant tunnelling.
If the layers are fabricated from the doped semiconductors, the thermal and
electric-field induced ionization of donor and acceptor impurities yields
additional charge carriers, which are non-uniformly distributed over the
nanostructure.
These charge carriers are sources of an additional electric field which leads
to the second component of the electron potential.
The theoretical description should take into account both the
components of the electron potential energy, whereas the density of the
additional charge carriers and its electrostatic potential should be calculated
by a self-consistent manner.
The present approach is based on the self-consistent solution of the quantum
kinetic equation for the Wigner distribution function and Poisson equation
for the electrostatic potential.

\begin{figure}
\begin {center}
\includegraphics[width=0.7 \textwidth]{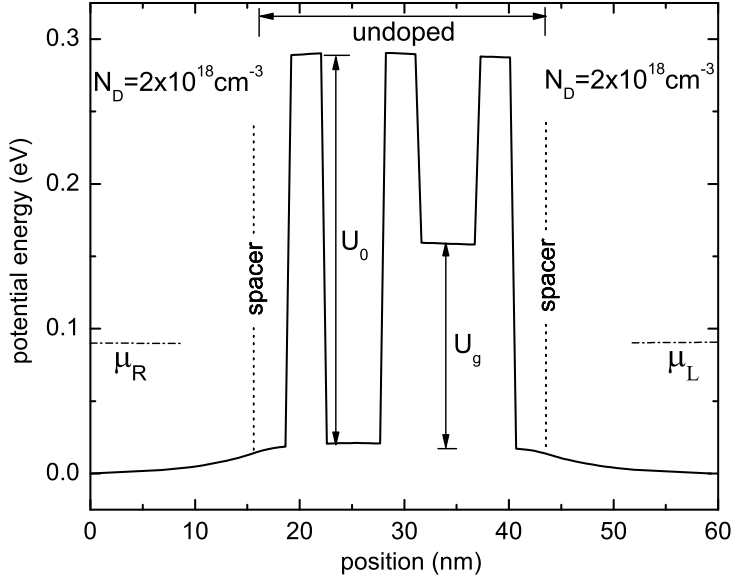}
\caption{Self-consistent potential energy profile in the TBRTD for zero bias
voltage.
The position is measured in the layer growth direction, $\mu_{L(R)}$ is the
electrochemical potential of the left (right) contact, $U_0$ is the height of
the potential barrier, and $U_g$ is the gate-voltage controlled shift of the
bottom of the right potential well.
The undoped region  consists of three Al$_{0.3}$Ga$_{0.7}$As
potential barriers and two GaAs potential wells
sandwiched between two spacer regions.}
\label{fig:1}
\end{center}
\end{figure}

In the quantum kinetic theory, the conduction band electrons are described by
the WDF.
If we take on the $x$ axis in the growth direction of the semiconductor layers
forming the TBRTD and assume the translational symmetry in the lateral
directions $(y,z)$ (cf. figure~\ref{fig:1}), we can obtain the basic electron
transport characteristics of the TBRTD from the Wigner distribution function
$\rho_w(x,k,t)$.
Neglecting the scattering processes and intervalley transitions in the
conduction band, we write the quantum kinetic equation in the
form~\cite{f:ineqsm, Ferry1999}
\begin{equation}
 \frac{\partial \rho_w(x,k,t)}{\partial t}
 + \frac{\hbar k}{m}\frac{\partial \rho_w(x,k,t)}{\partial x}
 = \frac{i}{2\pi \hbar}\int{dk^{\prime}}~U_w(x,k-k^{\prime})
 \rho_w(x, k^{\prime},t) \; ,
\label{eq:WEq}
\end{equation}
where $m$ is the conduction band effective mass and the integral kernel
$U_w(x,k-k^{\prime})$ represents the non-local potential energy defined as
\begin{equation}
 U_w(x,k-k^{\prime}) =
 \int{dx^{\prime}}~\big[U(x+x^{\prime}/2)-U(x-x^{\prime}/2)\big]
 \exp{\big[-i(k-k^{\prime})x^{\prime}\big]} \; ,
\label{eq:npot}
\end{equation}
where $U(x)$ is the total electrostatic potential energy of the electron.

In present work, we consider the steady-state solutions of~(\ref{eq:WEq}),
i.e., the solutions obtained under assumption
$\partial \rho_w(x,k,t)/\partial t =0$.
The potential energy $U(x)$ consists of the conduction band potential
$U_B(x)$ and the Hartree potential energy $U_H(x)$ of the electrons which
originates from the ionized donor impurities (the $n$-type doping is assumed).
Explicitly,
\begin{equation}
U(x)=U_B(x)+U_H(x).
\label{U}
\end{equation}
The conduction band component $U_B(x)$ can be written down in the form
\begin{equation}
U_B(x)=\sum_{i=1}^NU_i\Theta(x-x_i)\Theta(x_{i+1}-x) \; ,
\label{eq:cpot}
\end{equation}
where $N$ is the number of barrier regions, $x_i$ is the position barrier-well
(well-barrier) interface, i.e., $|x_{i+1}-x_i|$ determines the thickness of the
barrier region, $\Theta(x-x_i)$ is the Heaviside step function, and $U_i$ is the
height of the $i$-th barrier.

The Poisson equation for the Hartree potential energy takes on the form
\begin{equation}
\frac{d^2U_H(x)}{dx^2}=\frac{e^2}{\varepsilon_0\varepsilon}[N_D(x)-n(x)] \; ,
\label{eq:PEq}
\end{equation}
where $e$ is the elementary charge, $\varepsilon_0$ is the vacuum electric
permittivity, $\varepsilon$ is the relative static electric permittivity,
$N_D(x)$ is the concentration of the ionized donors, and $n(x)$ is the
electron density.

If we determine the WDF, we can find from its zero and first moments
the electron density and current density, respectively~\cite{f:ineqsm}.
In the explicit form,
\begin{equation}
n(x)=\frac{1}{2\pi}\int{dk} \rho_w(x,k)
\label{eq: ed}
\end{equation}
and
\begin{equation}
j(x)=\frac{e}{2\pi}\int{dk}\frac{\hbar k}{m} \rho_w(x,k).
\label{eq: fc}
\end{equation}

The quantum kinetic equation (\ref{eq:WEq}) and Poisson equation (\ref{eq:PEq})
should be solved simultaneously by a self-consistent
procedure~\cite{Kluksdahl1989}.
We assume the Dirichlet boundary conditions for the Poisson equation:
$U_H(0)=0$ and $U_H(L)=-eV_b$, where the bias voltage $V_b = V_R - V_L$ is
applied between the right ($R$) and left ($L$) electrodes separated by distance
$L$.
Moreover, for the WDF we take on  the open boundary conditions which have the
form~\cite{Frensley1990c1}
\begin{eqnarray}
\rho_w(0,k)\bigg|_{k>0}&=&f^{L}(k) \; , \\ \nonumber
\rho_w(L,k)\bigg|_{k<0}&=&f^{R}(k) \; ,
\label{eq:bc}
\end{eqnarray}
where $f^{L(R)}(k)$ is the supply function \cite{Ferry1999}, i.e., the
Fermi-Dirac distribution function integrated over the transverse momenta of
electrons in the left (right) reservoirs.
The supply function for the left (right) reservoir has the form
\begin{equation}
f^{L(R)}(k)=\frac{mk_BT}{\pi \hbar^2}
\ln{\bigg\{1+\exp{\bigg[-\frac{1}{k_BT}
\bigg(\frac{\hbar^2k^2}{2m}-\mu_{L(R)}\bigg)}\bigg]\bigg\}} \; ,
\label{eq:sf}
\end{equation}
where $T$ is the temperature and $\mu_{L(R)}$ is the electrochemical potential
of the left (right) reservoir.

In the present paper, we consider the TBRTD nanodevice composed of GaAs and
Al$_{0.3}$Ga$_{0.7}$As layers (figure~\ref{fig:1}).
The TBRTD nanodevice consists of the triple barrier region with two potential
wells, spacer layers, and contacts (cf. figure~\ref{fig:1}).
The contacts are made from the $n$-doped GaAs with homogeneous donor
concentration $N_D=2\times 10^{18}$cm$^{-3}$.
The calculations have been performed for the following parameters of the layers:
the thickness of each contact is equal to 17~nm, the thickness of each spacer
layer is 3~nm, the thickness of each potential barrier (well) is 3~nm (5~nm),
the height of the potential barrier $U_0 = 0.27$~eV, the total length of the
nanodevice is $L = 59$~nm.
In the calculations, we change energy $U_g$ of the bottom of the right potential
well which allows us to determine the effect of gate voltage applied to this
region.
We assume that the electrons are described by the conduction band effective mass
of GaAs, i.e., $m=0.667\,m_0$, where $m_0$ is the free electron mass.
Similarly, we take on the electric permittivity, $\varepsilon=12.9$,
and the lattice constant, $a=0.565$~nm for GaAs.

Temperature $T$ enters into the problem via supply function (\ref{eq:sf}),
which exhibits an almost unchanged shape in the interval $0 < T \leq 77$~K.
However, taking on the larger value of $T$ considerably accelerates the
calculations.
Therefore, in the present paper, we simulate the low-temperature properties of
the TBRTD putting $T$ = 77~K.

We have obtained the steady-state solutions of the Wigner-Poisson problem for
the TBRTD using the numerical scheme proposed by Biegel and
Plummer~\cite{Biegel1996} and Kim~\cite{Kim2007}.
In the first step, we find the steady-state Wigner function using the
time-independent form of~(\ref{eq:WEq}) with the potential energy given
by~(\ref{eq:cpot}) and calculate the electron density according
to~(\ref{eq: ed}).
In the next step, after inserting the electron density into~(\ref{eq:PEq})
we calculate the new potential energy profile.
Using this potential energy, we again solve the time-independent form
of~(\ref{eq:WEq}).
This procedure is repeated until the convergence is reached.
All the simulations were carried out using the computational grid with $N_x=106$
mesh points for the position $x$ and $N_k=72$ for the wave vector $k$.

\section{Results}\label{sec:results1}

In the gate-controlled TBRTD nanodevice, we have a possibility of designed
changing the energy levels of the electrons confined in the quantum wells.
In the considered TBRTD model, we can shift up and down the energy levels in the
right potential well by changing energy $U_g$ of the potential well bottom.
In a practical realization, $U_g$ can be changed by applying gate voltage $V_g$
to the nearby gate attached to the right potential well layer.
Then, $U_g = -e \alpha V_g$, where $\alpha$ is the voltage-to-energy conversion
factor which is characteristic to the nanodevice~\cite{Adamowski2006}.
Alternatively, $U_g$ can be changed in the technological process
by fabricating the right-potential well layer from Al$_x$Ga$_{1-x}$As with
$x < 0.3$.

\begin{figure}
\begin {center}
\includegraphics[width=0.7 \textwidth]{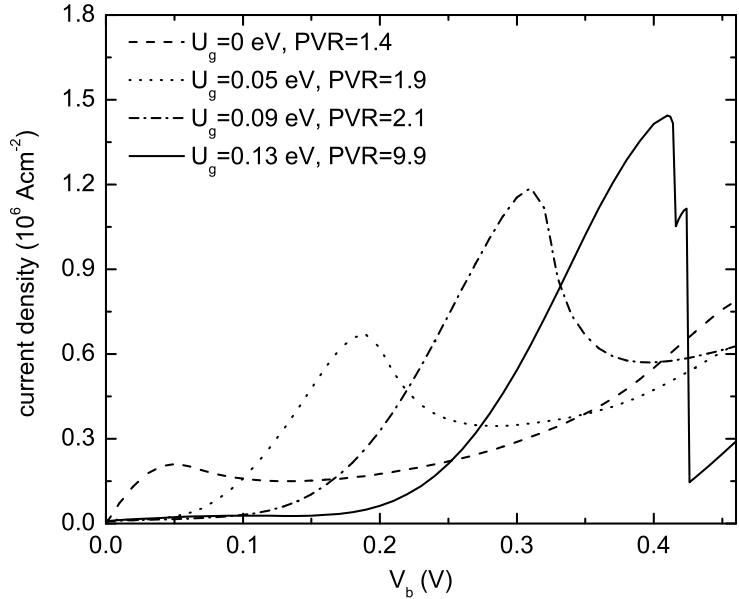}
\caption{Current--voltage characteristics and $PVR$ coefficients of the TBRTD
nanodevice calculated for different $U_g$.}
\label{fig:2}
\end{center}
\end{figure}

The resonant tunnelling devices are characterized by the Peak-to-Valley Ratio
($PVR$) which is defined as the ratio of the maximum (peak) current to the
minimum (valley) current on the current-voltage characteristics.
The results of our previous calculations~\cite{Wojcik2008} suggest the large
value of $PVR$ ($PVR = 27$) obtained for $U_g=0.1$~eV.
The present more realistic approach leads to smaller values of $PVR$
(figure~\ref{fig:2}).
The smaller values of the $PVR$ result from the considerable modification of the
potential profile due to the electrostatic interaction with the localized
electrons (cf. figure~\ref{fig:1}).
The maximum value of the $PVR$ obtained in the present calculations is 9.9.
We note, however, that this value of the $PVR$ is still larger than those
determined for the resonant-tunnelling diodes based on the GaAs
technology~\cite{Shewchuk1985}.

Figure~\ref{fig:2} shows the current-voltage characteristics of the TBRTD
nanodevice calculated  for different potential energy bottom $U_g$ of the right
potential well.
We observe that the increasing $U_g$ shifts the positions of the current peaks
towards the higher bias voltage.
This effect results from the increasing ground-state energy of the electron in
the double well potential.
Then, the condition of resonant tunnelling is satisfied at the higher bias
voltage.
We have found that for $U_g \leq 0.13$ eV the $PVR$ increases with the
increasing $U_g$.
At $U_g = 0.13$~eV the $PVR$ reaches the value 9.9 which is the maximal value of
the $PVR$ obtained in the present calculations.

\begin{figure}
\begin {center}
\includegraphics[width=0.7 \textwidth]{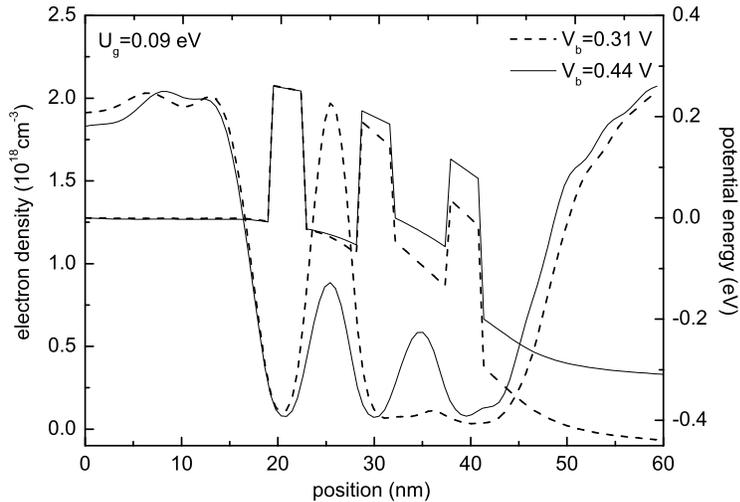}
\caption{Potential energy profile and electron density distribution for
$U_g=0.09$~eV and bias voltage $V_b$ = 0.31~V and 0.44~V that corresponds to the
current maximum and minimum, respectively.}
\label{fig:3}
\end{center}
\end{figure}

The obtained features of current-voltage characteristics can be explained if we
consider the self-consistent potential energy profiles and electron distribution
in the TBRTD.
Figure~\ref{fig:3} displays the results for $U_g$ = 0.09 eV.
We observe that the electron distribution for the bias corresponding to the
maximum of current ($V_b$ = 0.31 V) considerably differs from those for the
minimum of current ($V_b$ = 0.44 V).
The resonant tunnelling (current maximum) appears for the large electron
accumulation in the left potential well which results from the fact that -- in
this case -- the tunnelling current flows via the one-electron ground state
which is localized in this potential well.
For the minimal current the electrons occupy both quantum wells with the
comparable density.

\begin{figure}
\begin {center}
\includegraphics[width=0.7 \textwidth]{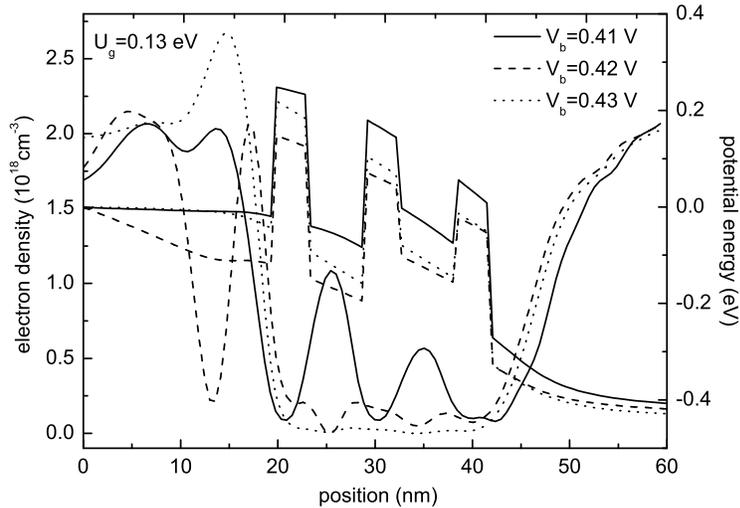}
\caption{Potential energy profile and electron density distribution for
$U_g=0.13$~eV and bias voltage $V_b$ = 0.41, 0.42, 0.43~V.}
\label{fig:4}
\end{center}
\end{figure}

A closer inspection of the current-voltage characteristics for $U_g=0.13$~eV
(figure~\ref{fig:2}) exhibits a notch in the regime of negative differential
resistance.
If the bias voltage becomes larger than 0.41 V, the current density stops to
decrease but starts to increase linearly with the increasing bias.
This increase occurs in the narrow bias voltage regime of the width estimated to
be 0.02 V.
This linear increase of the current is sometimes called the ''plateau-like''
behaviour \cite{plateau-endnote}.
For $V_b$ exceeding 0.43 V we again observe the negative differential resistance
(cf. figure~\ref{fig:2}).
This unusual property of the current-voltage characteristics results from the
formation of the potential well in the region of the left spacer, which
appears in the narrow bias voltage regime (figure~\ref{fig:4}).
The electron density distribution and potential profile in the bias voltage
regime 0.41 V $\leq V_b \leq$ 0.43 V are displayed in figure~\ref{fig:4}.
In this spacer-related potential well, the electrons can form the quasi-bound
states with the discrete energy levels.
The ground-state energy of the electron in the left spacer is slightly
lower than the energies of the quasi-bound states in the quantum wells which are
responsible for the resonant tunnelling through the TBRTD.
The spacer-related potential well leads to the accumulation of electrons flowing
from the left contact.
Because the difference between energy levels in the left contact and in the left
quantum well is very small, the conditions for resonant tunnelling are
approximately fulfilled.
In the bias voltage regime $0.41 V\leq V_b\leq0.43 V$ the electron
can flow from the left contact via the spacer and the triple barrier region
to the right contact.
Therefore, the notch on the current-voltage characteristics which appears in
this bias voltage regime, is a signature of the resonant tunnelling
through the spacer quasi-bound states.
If the bias voltage increases above $V_b$ = 0.43~V, the spacer-related potential
well becomes more shallow and finally disappears which rises the energy of the
electrons in the spacer.
This leads to the charge outflow from the left spacer
(cf. the curves for $V_b$ = 0.43 V in figure~\ref{fig:4}).
At higher bias voltage the resonant tunnelling is no longer responsible for the
electron transport which leads to an abrupt decline of the current.

The change of the current-voltage characteristics which appears for $U_g \geq
0.13$ eV, results from the considerable modification of the potential profile
which is visible if we compare the potential profile for $U_g$ = 0.13~eV
(figure~\ref{fig:4}) with those for $U_g$ = 0.09~eV (figure~\ref{fig:3}).
The results presented in figure~\ref{fig:3} show that the spacer-related
potential well almost disappears.
Therefore, there is no charge accumulation in the spacer region.
The plots presented in figure~\ref{fig:3} correspond to the usually observed
behaviour of current vs voltage in the RTD.

\begin{figure}
\begin {center}
\includegraphics[width=0.7\textwidth]{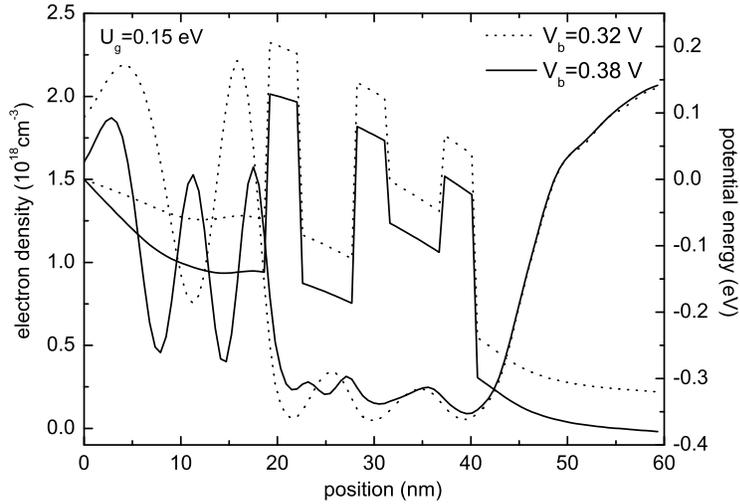}
\caption{Potential energy profile and electron density distribution for
$U_g=0.15$~eV and for bias voltages $V_b$ = 0.32 V and 0.38~V,
which correspond to the two linear-growth regimes in the current-voltage
characteristics depicted in figure~\ref{fig:6} (c).}
\label{fig:5}
\end{center}
\end{figure}

\begin{figure}
\begin {center}
\includegraphics[width=0.49 \textwidth]{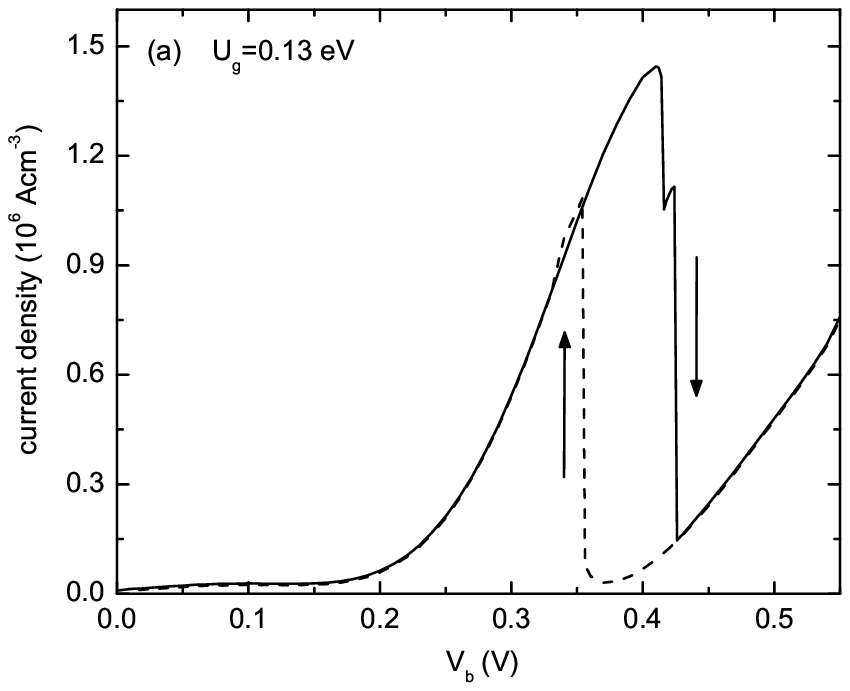}
\includegraphics[width=0.49 \textwidth]{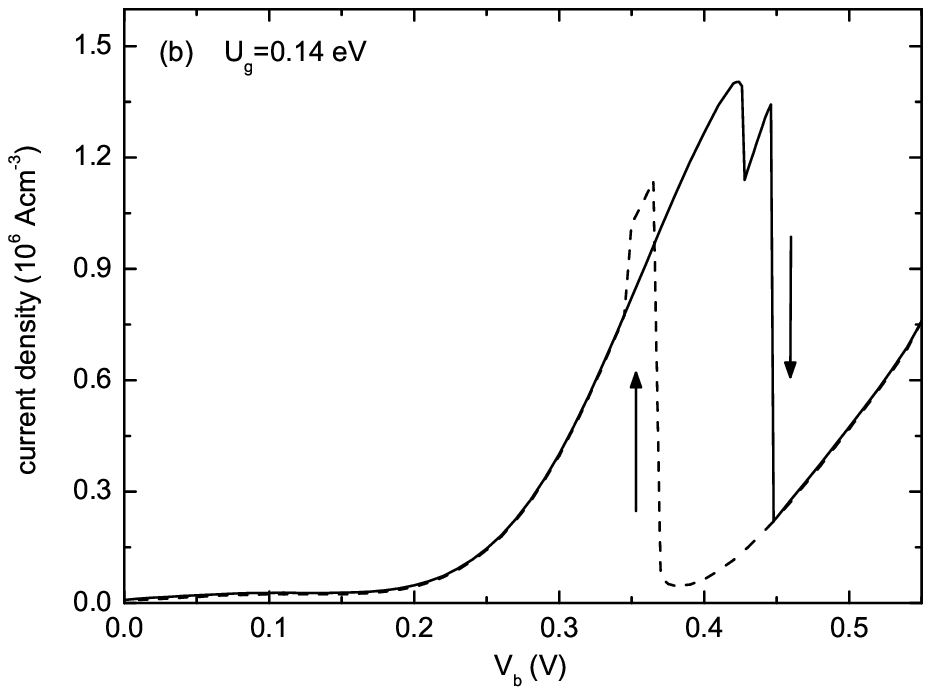}
\includegraphics[width=0.49 \textwidth]{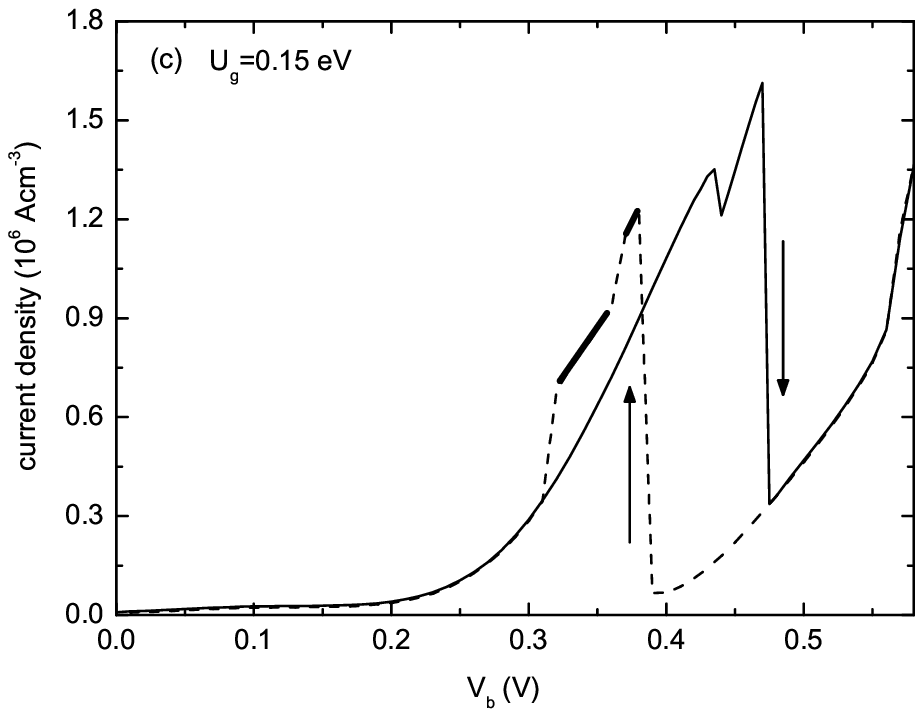}
\caption{Current-voltage characteristics of the TBRTD for (a) $U_g$ = 0.13 eV,
(b) $U_g$ = 0.14 eV, and (c) $U_g$ = 0.15 eV.
Solid (dashed) curves with arrows down (up) correspond to the results obtained
for increasing (decreasing) bias voltage.}
\label{fig:6}
\end{center}
\end{figure}

If $U_g$ exceeds 0.13 eV, the negative-differential resistance behaviour is
further modified (figure~\ref{fig:6}): the notch becomes more distinct and the
width of the linear-increase region is extended.
At $U_g$ = 0.15 eV, the notch dominates
in the current-voltage characteristics and the current maximum has the form of
the sharp peak reached after the linear-increase behaviour
[figure~\ref{fig:6}(c)].
The linear current versus bias voltage behaviour has been obtained
for the increasing as well decreasing bias voltage.
For the decreasing bias voltage and $U_g=0.15$~eV we have obtained
the following two linear current-versus-voltage regimes: (i) 0.32 V $\leq V_b
\leq$ 0.36 V, and (ii) 0.37 V $\leq V_b \leq$ 0.38 V [cf. bold lines
in \ref{fig:6}(c)].
The potential profile and electron density for these two regimes are depicted in
figure~\ref{fig:5}.
The difference in the electron distribution (figure~\ref{fig:5}) for these two
regimes results from the formation of quasi-bound states in the spacer layer.
This in turn leads to the appearance of distinct regimes (i) and (ii).
In regime (i), the transmission via the spacer-related ground state dominates in
the resonant tunnelling, while in regime (ii), we deal with the tunnelling via
the first excited state.
We have checked the above conclusions by a direct numerical calculation
of the quasi-bound state energy levels.

Additionally, we have found that  the current-voltage characteristics
exhibit the bistability (figure~\ref{fig:6}).
Figures~\ref{fig:6}(a,~b,~c) show that the current-voltage characteristics
obtained for the increasing bias (solid curves) differ from those for the
decreasing bias.
The largest difference is observed in the region of maximal current.

\begin{figure}
\begin {center}
\includegraphics[width=0.7 \textwidth]{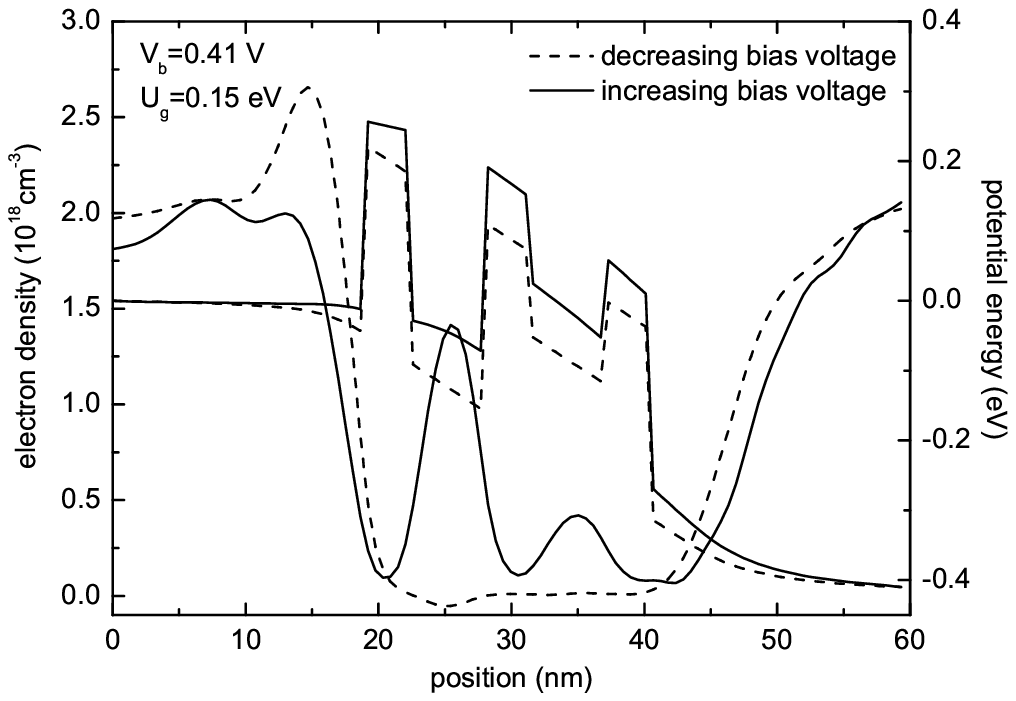}
\caption{Potential energy profile and electron density distribution in the TBRTD
for $U_g$ = 0.15 eV and for bias $V_b$ = 0.41 V which corresponds to the
bistability region in figure \ref{fig:6}.
Solid (dashed) curves show the results obtained for the increasing
(decreasing) bias voltage.}
\label{fig:7}
\end{center}
\end{figure}

The bistability in TBRTD current-voltage characteristics can be
explained by either the mathematical or physical arguments.
If we consider the mathematical properties of the self-consistent
Wigner-Poisson problem, we observe its nonlinearity.
The solutions of the nonlinear problem are very sensitive to the initial
conditions, therefore, they depend on the direction of the bias voltage change.
Physical there are two orgins of bistability.
First of them is the feedback between the charge accumulated in quantum wells
and the flowing current~\cite{Goldman87, Rahman1990},
and the second is a current oscillation~\cite{Sollner1987}.
The bias voltage $V_b=0.41$ V corresponds to one of the two bistability points
in the current-voltage characteristics as it is shown in figure~\ref{fig:7}.
We observe that both the potential energy and electron density are considerably
different in the cases of increasing and decreasing bias voltage.
The increasing bias causes that the electrons are localized in both
the potential wells, however, the electron localization in the left quantum well
is stronger (cf. solid curves in figure~\ref{fig:7}).
The electron localization results from the formation of the quasi-bound states
in both the potential well regions, which -- at the resonant conditions -- leads
to the increase of the current.
This type of behaviour is typical for asymmetric potential profile.
On the contrary, the decreasing bias leads to the accumulation of electrons in
the left spacer (near the left potential barrier).
In this case, the electron density is almost zero within the potential wells
which results in the considerable reduction of the maximal current
[cf. figure~\ref{fig:6}(c)].

\section{Conclusions and Summary}\label{sec:concl}

In the present paper, we have studied the transport properties of electrons in
the triple-barrier resonant tunnelling diode.
We have solved the Wigner-Poisson problem by the self-consistent procedure and
determined the current-voltage characteristics for different gate voltage
applied to the right potential well region.
We have found the fairly large value ($\sim$10) of the peak-to-valley ratio.
The enlargement of the peak-to-valley ratio over that observed for the
double-barrier resonant tunnelling diodes is caused by appearance of the linear
regime within the negative differential resistance region that is produced by
the join effect of the additional potential barrier and potential well.
The additional potential barrier leads to the accumulation of electrons in the
left spacer, while the second potential well leads to the formation of
quasi-bound states in this potential well.
For properly chosen nanodevice parameters the resonance conditions are satisfied
for the quasi-bound states appearing in the three potential well regions,
namely, the left spacer and two potential wells.
This leads to a considerable increase of the resonant current peak.

We have also found interesting properties of the current-voltage characteristics
of the TBRTD: the change of the negative differential resistance into the
positive one (the occurrence of the notch in the
regime of the linear dependence of the current on the bias voltage)
and the bistability.
We have explained these properties by analyzing the corresponding changes of the
potential profile and electron density  which are controlled by the gate
potential.
In the present study, we focus on the influence of the self-consistent
Hartree potential on the transport properties of the TBRTD, therefore, we
neglect the electron-phonon scattering.
The bistable behaviour found in the current-voltage characteristics of the TBRTD
requires a further study that should include the time dependence of the current.

In summary, the current-voltage characteristics of the triple-barrier resonant
tunnelling diode shows a variety of interesting properties  that can be
effectively tuned by applying the gate voltage.

\section*{Acknowledgements}\label{sec:aments}

This paper has been supported by the Foundation for Polish Science MPD Programme
co-financed by the EU European Regional Development Fund and the Polish Ministry
of Science and Higher Education Programme ''International scientists mobility
support''.

\end{document}